\documentclass[amsmath,11pt]{article}
\usepackage{graphicx}

\date{\empty}

\begin{document}

\title{\bf Melvin's ``magnetic universe'', the role of the magnetic tension and the implications for gravitational collapse}

\author{Christos G. Tsagas${}^{1,2}$ and Panagiotis Mavrogiannis${}^1$\\ ${}^1${\small Section of Astrophysics, Astronomy and Mechanics, Department of Physics}\\ {\small Aristotle University of Thessaloniki, Thessaloniki 54124, Greece}\\ ${}^2${\small Clare Hall, University of Cambridge, Herschel Road, Cambridge CB3 9AL, UK}}

\maketitle

\begin{abstract}
Electromagnetism appears to have the potential to alter the fate of relativistic gravitational implosion. Coulomb forces, for example, can in principle prevent the ultimate collapse of charged matter to a singularity. Also, half a century ago, Mael Melvin noted that magnetic forcelines do not collapse under their own gravity, no matter how strong the latter may be. Instead, magnetic fields seemed capable of stabilising themselves against gravitational self-contraction. Intrigued, Melvin wondered whether magnetism could also support against the realistic collapse of charged matter. Here, we look back into these issues by means of two complementary scenarios. The first reconsiders Melvin's highly idealised ``magnetic universe'' and reaches the same conclusion, namely that magnetic fields do not self-gravitate. The second is a realistic collapse scenario, like that of magnetised massive compact stars. Looking for answers to Melvin's question, we find that magnetic fields do show a generic tendency to support against relativistic gravitational implosion. The magnetic effect in question involves the curvature of the host space, it grows stronger as the collapse progresses and the triggering agent is the field's tension. In other words, the reason is the elasticity of the field lines and their inherent tendency to react against anything that distorts them from equilibrium. In this case, the distorting agent is the ever increasing curvature of the collapsing host space.
\end{abstract}

\section{Introduction}\label{sI}
Posing idealised physical questions allows one to push the laws of physics to the extreme and thus expect to gain new insight into the way these laws work~\cite{T}. This was what Mael Melvin did half a century ago, when he considered the contraction and collapse of a cylindrical tube of source-free magnetic lines, resting in an otherwise empty and static space, under their own gravity~\cite{M1,M2}. He found that, irrespective of how strong the magnetic gravitational field was, the forcelines never collapsed. Instead, the magnetic ($B$) field showed an unexpected ability to stabilise itself against gravitational self-contraction. Intrigued by his findings, Melvin wondered what they might mean for realistic gravitational collapse. In particular, he posed the question as to whether the magnetic ability to stabilise itself against self-contraction, also reflected some underlying natural tendency of the $B$-field to resist the collapse of charged matter as well~\cite{M2}. Decades later, soon after the turn of the millennium, independent studies suggested that magnetic fields might actually have such an inherent tendency to resist relativistic gravitational implosion~\cite{T1,T2}.

These are not the only occasions where the involvement of electromagnetism was found to affect and eventually alter the traditionally expected outcome of gravitational implosion. Studies of charged collapse, for example, have opened the possibility that repulsive Coulomb forces could (in principle) lead to a bounce and thus prevent the formation of singularities (e.g.~see~\cite{Nov}-\cite{KB}). With this in mind, we take an alternative look into Melvin's work and its potential implications by employing two complementary scenarios. The first (see scenario (a) in \S~\ref{sTSs} below) reconsiders the gravitational self-contraction and collapse of source-free magnetic forcelines resting in otherwise empty space, thus reproducing Melvin's highly idealised magnetic universe. Although our approach is different, since we look at the convergence/divergence of the (spacelike) magnetic forcelines by means of the Raychaudhuri equation, we arrive at the same conclusion. We find that the magnetic field lines do not converge under their own gravity no matter how strong the latter is. Moreover, we identify the magnetic tension as the reason behind this unconventional magnetic behaviour. The latter cancels out the contribution of the field's energy density to the local curvature and in so doing prevents the forcelines from converging. In a sense, the magnetic tension ensures that $B$-fields cannot ``feel'' their own gravitational pull.

Our second scenario goes beyond Melvin's idealised model. There, we consider the realistic gravitational collapse of highly conductive magnetised matter, like that occurring in massive compact stars for example. In this effort, we are also motivated by the question Melvin posed but left unanswered, namely whether magnetism can support against the gravitational collapse of charged media. We employ again the Raychaudhuri equation, though this time it is applied to the timelike congruence of the matter particles (see scenario (b) in \S~\ref{sTSs}). The latter are tightly coupled to the magnetic forcelines, given the high electrical conductivity of the collapsing fluid. Our study shows that $B$-fields have a generic tendency to resist the collapse, which is purely relativistic in nature. More specifically, the magnetic presence naturally gives rise to supporting magneto-curvature stresses. These are triggered by the magnetic tension, namely by the elasticity of field lines, and reflect their inherent tendency to remain ``straight'' and to react against any agent that distorts them from equilibrium. In this case, the line deformation is caused by the curvature of the host space and, as a result, the supporting tension stresses grow stronger as the gravitational contraction progresses and the curvature distortion increases. Put another way, although the tension stresses are triggered by the magnetic presence, their resisting effect is driven by the increasing curvature of the collapsing host space. Therefore, in principle at least, even a relatively weak $B$-field could lead to a disproportionately strong magneto-curvature support during the advanced stages of the collapse. In a sense, it appears as though the elastic properties of the magnetic forcelines have been transferred into the fabric of the host space, which seems to act like a spring under tension.

\section{The Raychaudhuri equations}\label{sREs}
Theoretical studies of gravitational collapse typically employ the Raychaudhuri equation, which has been at the centre of the singularity theorems (e.g.~see~\cite{HE}-\cite{Po}). Although Raychaudhuri's formula is usually applied to the timelike worldlines of real (or hypothetical) observers, it can be used (and it has been used) to study the ``kinematics'' of spacelike and null curves as well (e.g.~see~\cite{KS,AV}). The magnetic forcelines, as seen by observers moving along their own timelike paths, are examples of such spacelike curves. Technically speaking, if $u_a$ (with $u_au^a=-1$) is the timelike 4-velocity of a family of observers, the magnetic field vector ($B_a$) seen by them satisfies the constraint $B_au^a=0$. The observers' 4-velocity vector, together with the associated projection tensor $h_{ab}=g_{ab}+u_au_b$ (where $g_{ab}$ is the spacetime metric and $h_{ab}u^b=0$) introduces an 1+3 decomposition of the spacetime into time and 3-space. Similarly, if $n_a$ is a unit spacelike vector along the direction of the magnetic lines (with $B_a=Bn_a$), the tensor $\tilde{h}_{ab}= h_{ab}-n_an_b$ projects onto the 2-dimensional surface normal to $n_a$, namely $\tilde{h}_{ab}n^b=0$, thus achieving an 1+2 decomposition of the 3-space and leading to an overall 1+1+2 splitting of the whole spacetime.\footnote{Following our 1+1+2 spacetime splitting, we will use overdots to indicate time derivatives, namely differentiation along $u_a$ (i.e.~${}^{\cdot}=u^a\nabla_a$) and primes to denote differentiation along $n_a$ (i.e.~${}^{\prime}=n^a{\rm D}_a$). Note that $\nabla_a$ is the 4-dimensional covariant derivative operator, while ${\rm D}_a=h_a{}^b\nabla_b$ and $\tilde{\rm D}_a=\tilde{h}_a{}^b{\rm D}_b$ are its 3-dimensional and 2-dimensional counterparts respectively (see~\cite{TCM}-\cite{GT} for further details and references).}

By construction, the $u_a$ and the $n_a$ fields are tangent to a family of timelike worldlines and to a congruence of spacelike curves respectively. Then, the variables $\Theta={\rm D}^au_a$ and $\tilde{\Theta}=\tilde{\rm D}^an_a$ respectively define the 3-dimensional volume expansion/contraction scalar and the 2-dimensional area scalar (spatial divergence) of the aforementioned two groups of curves. Consequently, positive values for $\Theta$ and $\tilde{\Theta}$ imply that the tangent curves move apart (i.e.~expansion), whereas in the opposite case we have contraction and the curves are approaching each other. The kinematics of the expansion/contraction are monitored by the associated Raychaudhuri equations. In particular, the convergence or divergence of the timelike worldlines tangent to the $u_a$-field follows from (e.g.~see~\cite{TCM,EMM})
\begin{equation}
\dot{\Theta}= -{1\over3}\,\Theta^2- R_{ab}u^au^b- 2\left(\sigma^2-\omega^2\right)+ {\rm D}^a\dot{u}_a+ \dot{u}^a\dot{u}_a\,.  \label{tmRay1}
\end{equation}
On the other hand, the converegence/divergence of the spacelike curves tangent to the $n_a$-field (i.e.~of the magnetic forcelines) is governed by~\cite{GT}\footnote{Expression (\ref{tmRay1}) monitors the convergence/divergence of the $u_a$-field in time (i.e.~along the $u_a$-field), while Eq.~(\ref{spRay1}) does the same for the $n_a$-congruence along a given spatial direction (i.e.~along the spacelike vector $n_a$).}
\begin{equation}
\tilde{\Theta}^{\prime}= -{1\over2}\,\tilde{\Theta}^2- \mathcal{R}_{ab}n^an^b- 2\left(\tilde{\sigma}^2-\tilde{\omega}^2\right)+ \tilde{\rm D}^an_a^{\prime}- n^{\prime\,a}n_a^{\prime}\,. \label{spRay1}
\end{equation}
Note that $R_{ab}$ and $\mathcal{R}_{ab}$ are respectively the 4-D (spacetime) and the 3-D (spatial) Ricci tensors. Also, $\sigma^2={\rm D}_{\langle b}u_{a\rangle}{\rm D}^{\langle b}u^{a\rangle}/2$ and $\tilde{\sigma}^2=\tilde{\rm D}_{\langle b}n_{a\rangle}\tilde{\rm D}^{\langle b}n^{a\rangle}/2$ are the (squared) magnitudes of the shear tensors associated with the $u_a$ and the $n_a$ fields respectively, while $\omega^2={\rm D}_{[b}u_{a]}{\rm D}^{[b}u^{a]}/2$ and $\tilde{\omega}^2=\tilde{\rm D}_{[b}n_{a]}\tilde{\rm D}^{[b}n^{a]}/2$ are the magnitudes of the corresponding vorticity tensors. Finally, $\dot{u}_a=u^b\nabla_bu_a$ and $n_a^{\prime}=n^b{\rm D}_bn_a$ are the directional-derivative vectors ($\dot{u}_a$ is the 4-acceleration) associated with the aforementioned two families of curves. When $\dot{u}_a=0$ the observers' worldlines are timelike geodesics, whereas $n_a^{\prime}=0$ implies that the $n_a$-field is tangent to a congruence of spacelike geodesics. The very close analogy between expressions (\ref{tmRay1}) and (\ref{spRay1}) is profound, with every term on the right-hand side of (\ref{spRay1}) having a direct analogue in Eq.~(\ref{tmRay1}). In what follows, we will use the above two versions of the Raychaudhuri equation to argue that:
\begin{itemize}
\item (a) The magnetic lines do not self-gravitate, because they do not ``feel'' their own gravity no matter how strong the latter may be (in agreement with Melvin).\\
\item (b) In an external gravitational field, the magnetic field develops naturally curvature related stresses that resist gravitational contraction (in response to Melvin).
\end{itemize}
In both cases, we will show that the agent responsible for such a unique and unconventional behaviour is the magnetic tension.

\section{The two scenarios}\label{sTSs}
\textbf{Scenario~(a):} For our purposes, the key difference between Eqs.~(\ref{tmRay1}) and (\ref{spRay1}) is in their curvature terms. In particular, the former relation is affected by the 4-Ricci curvature of the whole spacetime, while the latter contains the 3-Ricci tensor of the spatial hypersurfaces. These are
\begin{eqnarray}
R_{ab}= {1\over2}\,\kappa\left(\rho+3p\right)u_au_b+ {1\over2}\,\kappa\left(\rho-p\right)h_{ab}+ 2\kappa q_{(a}u_{b)}+ \kappa\pi_{ab}  \label{Rab}
\end{eqnarray}
and
\begin{eqnarray}
\mathcal{R}_{ab}&=& {1\over3}\,\mathcal{R}h_{ab}+ {1\over2}\,\kappa\pi_{ab}+ E_{ab}- {1\over3}\,\Theta\left(\sigma_{ab}+\omega_{ab}\right)+ \sigma_{c\langle a}\sigma^c{}_{b\rangle}- \omega_{c\langle a}\omega^c{}_{b\rangle} \nonumber\\ &&+2\sigma_{c[a}\omega^c{}_{b]}\,,  \label{cRab}
\end{eqnarray}
respectively (e.g.~see~\cite{TCM,BMT}). In the above $\kappa=8\pi G$ is the gravitational constant, $\rho$ is the energy density of the (total) matter, $p$ is its isotropic pressure $q_a$ the associated energy flux and $\pi_{ab}$ the anisotropic pressure (i.e.~the viscosity), with the last two quantities vanishing in the case of a perfect fluid. Also, $\mathcal{R}=2(\kappa\rho-\Theta^2/3+ \sigma^2-\omega^2)$ is the 3-Ricci scalar (describing the mean curvature of the spatial sections) and $E_{ab}$ is the electric Weyl tensor (associated with tidal forces and gravitational waves). Note that $q_a$, $\pi_{ab}$ and $E_{ab}$ are spacelike, with the last two also being symmetric and trace-free. A key (and also very well known) difference between expressions (\ref{Rab}) and (\ref{cRab}) is that the former contains only contributions from the local matter, while the latter carries additional effects and in particular those of the long-range gravitational field, namely of the Weyl curvature (see also \S~\ref{sD} below for additional related comments).

Magnetic fields correspond to an effective imperfect fluid with energy density $\rho_{_B}=B^2/2$ (where $B^2=B_aB^a$), isotropic pressure $p_{_B}=B^2/6$ and anisotropic pressure $\pi_{ab}=\Pi_{ab}=(B^2/3)h_{ab}-B_aB_b$ (e.g.~see~\cite{BMT} and references therein). This tensor also carries the effects of the field's tension, which is manifested as negative pressure exerted along the direction of the magnetic lines. Indeed, one can easily show that $\Pi_{ab}$ has a negative eigenvalue (equal to $-2B^2/3$) along the direction of the $B$-field.

Let us now apply expressions (\ref{spRay1}) and (\ref{cRab}) to Melvin's ``magnetic universe''. In order to test the behaviour of the $B$-field under its own gravity, Melvin considered a cylindrical tube of infinitely long, parallel (i.e.~irrotational and shear-free spacelike geodesics) sourceless forcelines, resting in an otherwise empty and static space~\cite{M1,M2}. In such an environment, where there is nothing whatsoever to affect (or be affected) by the magnetic field, Eqs.~(\ref{spRay1}) and (\ref{cRab}) reduce to
\begin{equation}
\tilde{\Theta}^{\prime}= -{1\over2}\,\tilde{\Theta}^2- \mathcal{R}_{ab}n^an^b  \label{MspRay1}
\end{equation}
and
\begin{eqnarray}
\mathcal{R}_{ab}&=& {1\over3}\,\kappa B^2h_{ab}+ {1\over2}\,\kappa\Pi_{ab}\,,  \label{McRab}
\end{eqnarray}
respectively~\cite{GT}. Note the term $\mathcal{R}_{ab}n^an^b$ on the right-hand side of (\ref{MspRay1}), which measures the 3-curvature along the direction of the $B$-field and essentially describes distortions in the shape of the magnetic forcelines caused by the curvature of the host space. Contracting Eq.~(\ref{McRab}) twice along $n_a$ gives $\mathcal{R}_{ab}n^an^b=0$ irrespective of the field's strength, given that $h_{ab}n^an^b=n_an^a=1$ and $\Pi_{ab}n^an^b=-2B^2/3$ (see above). This null result is a direct consequence of the magnetic tension, which cancels out exactly the contribution of the magnetic energy density in Eq.~(\ref{McRab}) and in so doing demonstrates the natural preference of the field lines to remain straight.

\begin{figure}
\centering
\includegraphics[width=0.7\columnwidth]{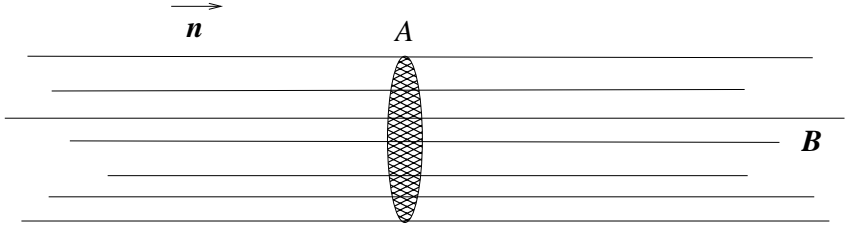}
\caption{Magnetic lines lying parallel in empty 3-space along the unit vector $\vec{n}$, with 2-D cross-sectional area $A$ (so that $A^{\prime}/A=\tilde{\Theta}$). The forcelines will not self-gravitate and converge (along their own direction), no matter how strong their gravitational field is. Instead, they remain at rest, due to their tension properties, provided $\tilde{\Theta}_0=0$ and no external agent intervenes.}  \label{fig:Blines}
\end{figure}

The vanishing of the curvature term on the right-hand side of (\ref{MspRay1}) means that the convergence/divergence of the forcelines (along their own direction) proceeds independently of the field's own gravity. In a sense, the magnetic lines do not ``feel'' their own gravitational field, no matter how strong the latter may be. As Melvin also claimed, the $B$-field does not self-gravitate.\footnote{The absence of a magnetic contribution to the scalar $\mathcal{R}_{ab}n^an^b$ is independent of the presence of other matter and of the strength of the $B$-field. If there are additional matter sources around, the magnetic forcelines will generally ``feel'' the gravitational pull of those sources, but never their own gravity (see scenario (b)).} Then, the convergence/divergence of the magnetic lines is fully determined by the initial conditions. More specifically, when $\mathcal{R}_{ab}n^an^b=0$, Eq.~(\ref{MspRay1}) gives
\begin{equation}
\tilde{\Theta}=\tilde{\Theta}(s)= {2\tilde{\Theta}_0\over2+\tilde{\Theta}_0s}\,,  \label{MtTheta}
\end{equation}
where $s$ is the proper length in the direction of the $B$-field. Consequently, when $\tilde{\Theta}_0<0$, the forcelines converge to form caustics within finite proper length (i.e.~$\tilde{\Theta}\rightarrow-\infty$ as $s\rightarrow -2/\tilde{\Theta}_0$). In the opposite case (when $\tilde{\Theta}_0>0$), the lines keep diverging. Most interestingly, however, magnetic forcelines that happen to be at rest (i.e.~with $\tilde{\Theta}_0=0)$ will remain so indefinitely (see Fig.~\ref{fig:Blines}), unless an external agent interferes (see scenario (b) next). The latter case profoundly demonstrates the natural tendency of the field lines to remain straight/stable and also reveals that the responsible agent is their tension, the negative pressure of which cancels out the gravitational pull of the field's own energy density.

The pivotal role played by the magnetic tension in arriving at the above results should be noted and emphasised. If there was no tension, namely if $\Pi_{ab}n^an^b\geq0$, Eq.~(\ref{McRab}) would have led to $\mathcal{R}_{ab}n^an^b\geq \kappa B^2/3$. This in turn would have meant $\tilde{\Theta}^{\prime}<0$, even when $\tilde{\Theta}=0$ initially (see expression (\ref{MspRay1})). In such a case, the magnetic lines would have converged and focused, under their own gravity alone, within finite proper length.

In summary, our first scenario allowed for an alternative approach to Melvin's idealised magnetic universe, where the field lines form a cylindrical tube resting in empty space~\cite{M1,M2}. We have recovered Melvin's conclusion on the one hand, while on the other we have revealed the key role of the field's tension as the stability agent. Despite the idealised nature of the model, the fact that the tension cancels out the magnetic energy input to $\mathcal{R}_{ab}n^an^b$ remains and holds irrespective of the presence of matter, or of the magnetic strength. Next, we will use relativistic magnetohydrodynamics (MHD) to consider the realistic collapse of a self-gravitating magnetised fluid. In the process, we will also try to answer Melvin's question as to whether the $B$-field can support against the relativistic implosion of charged matter~\cite{M2}.\\

\textbf{Scenario~(b):} The magnetic tension reflects the elasticity of the field lines and their tendency to react against any agent that distorts them from equilibrium~\cite{Pa,M}. In scenario (a), we considered the idealised situation of magnetic forcelines self-gravitating in an otherwise empty space. There, there was nothing to affect or to be affected by the magnetic presence. Here, we will embed these lines into the gravitational field of a highly conductive charged medium. In such an environment, which defines the ideal-MHD limit, Ohm's law guarantees that the electric field vanishes and the currents keep its magnetic counterpart frozen into the fluid~\cite{Pa}-\cite{M}. Then, Maxwell's formulae (Faraday's law in particular) ensure that the magnetic strength is enhanced/depleted by the compression/expansion of the host medium (see~\S~5.1 in~\cite{BMT} for the form of Maxwell's equations at the MHD limit).

Magnetic-line deformations are usually caused by electrically charged particles, but in the context of general relativity spacetime curvature (i.e.~gravity in Einstein's interpretation) can also act as a deforming agent (e.g.~see~\cite{NOO}-\cite{T3}). When this happens, the field's tension gives rise to magneto-curvature stresses that react against the cause of their distortion, namely against gravity itself. To demonstrate all this, assume that the host spacetime is neither static nor empty, but dynamical and filled with highly conductive charged matter. In such an environment, the magnetic lines are no longer infinite, but closed according to Gauss's law (i.e.~${\rm D}^aB_a=0$) and standard MHD. Then, the process of gravitational contraction is monitored by the usual form of the Raychaudhuri equation (see expression (\ref{tmRay1}) here, as well as~\cite{W}). The latter is applied to the timelike worldlines of the matter particles, instead of the spacelike magnetic forcelines of scenario (a). Put another way, in scenario (b), it is the highly conductive matter that collapses and in the process pulls the frozen-in magnetic field along with it.

In the magnetic presence, the worldlines of the charged particles are no longer timelike geodesics. This is a crucial difference between scenario (b) and the typical (non-magnetised) models of gravitational collapse, where matter ``moves'' along geodesic worldlines under the effect of gravity alone (see~\cite{W,Po} and also footnote~6 here). Note that the non-geodesic nature of timelike worldlines is guaranteed by the nonzero 4-acceleration vector in Raychaudhuri's formula ($\dot{u}_a$ -- see Eq.~(\ref{tmRay1}) in \S~\ref{sREs}), which in turn manifests the presence of non-gravitational forces. This is a crucial common feature between this scenario and those claiming that Coulomb forces can prevent the ultimate collapse of charged matter and the formation of singularities (see~\cite{Nov}-\cite{KB}). In our case, however, it is not the Coulomb but the magnetic Lorentz force that plays the pivotal role. At the ideal MHD limit, the Lorentz force satisfies the momentum conservation law
\begin{eqnarray}
\left(\rho+p+B^2\right)\dot{u}_a= -{\rm D}_ap- {1\over2}\,{\rm D}_aB^2+ B^b{\rm D}_bB_a+ \dot{u}^bB_bB_a\,,  \label{MHD-NS}
\end{eqnarray}
where the second and third terms on the right-hand side split the magnetic Lorentz force into its pressure and tension parts respectively~\cite{TCM,EMM}. To proceed, let us suppose that both the matter and the magnetic energies have nearly uniform distribution (i.e.~assume that ${\rm D}_a\rho\simeq0\simeq{\rm D}_ap\simeq{\rm D}_aB^2$). Then, to leading approximation, we may bypass the effects from any inhomogeneities may exist in the fluid and the magnetic energy densities.\footnote{Addressing the fully nonlinear case remains an intractable task, even by numerical methods, since it requires solving the complete set of the Einstein-Maxwell equations. Simplifying steps are therefore necessary and the homogeneity assumption is a standard practise, also adopted by the singularity theorems~\cite{W,Po}.} On the other hand, we allow for $B^b{\rm D}_bB_a\neq0$ and therefore incorporate effects caused by distortions in the shape of the magnetic forcelines. This enables us to focus on the magnetic tension (see decomposition (\ref{MHD-NS}) above), which is at the centre of this study.

Some clarifying notes are due here. Strictly speaking, hereafter, our analysis refers to matter fields with spatially homogeneous energy-density distribution. In practice, however, our results also apply to nearly homonegeous environments, where the matter fields involved have been sufficiently ``randomised''. It should also be noted that spatial homogeneity is not an ad hoc assumption of ours, but a rather common approximation. In fact, exact spatial homogeneity is a standard feature/prerequisite in all typical singularity theorems (e.g.~see~\cite{W,Po} and also footnotes~4 and~6 here). Recall that these theorems consider geodesics worldlines, a condition that no longer holds when matter and pressure inhomogeneities are present. Moreover, assuming a homogeneous (or nearly homogeneous) matter distribution does not compromise the core of our argument, since gradients in the fluid and in the magnetic density distribution are known to inhibit gravitational contraction, even within Newtonian physics. Here, on the other hand, we focus upon the purely general relativistic effects. On these grounds, the 3-divergence of Eq.~(\ref{MHD-NS}) combines with the 3-Ricci identities and with Maxwell's formulae to give~\cite{T1,T2}
\begin{equation}
{\rm D}^a\dot{u}_a= c^2_a\,\mathcal{R}_{ab}n^an^b+  2\left(\sigma^2_{_B}-\omega^2_{_B}\right)\,,  \label{Ddu}
\end{equation}
where $c^2_a=B^2/(\rho+p+B^2)$ is the Alfv\'en speed. Note that the 3-curvature term $\mathcal{R}_{an}n^an^b$ is nonzero now, due to the presence of matter, though still without any magnetic input. Also, the scalars $\sigma^2_{_B}={\rm D}_{\langle b}B_{a\rangle}{\rm D}^{\langle b}B^{a\rangle}/2(\rho+p+B^2)$ and $\omega^2_{_B}={\rm D}_{[b}B_{a]}{\rm D}^{[b}B^{a]}/2(\rho+p+B^2)$ are the magnetic analogues of the shear and the vorticity respectively.

All the terms seen on the right-hand side of Eq.~(\ref{Ddu}) result from the magnetic tension and reflect the ``elastic'' reaction of the field lines to distortions triggered by spatial curvature, shear and rotation respectively (see also expression (\ref{MHDtmRay1}) and related discussion below). The most important for our purposes is the first term, which is a (purely relativistic) magneto-curvature tension stress, since it contains contributions from the $B$-field and the 3-Ricci tensor (contracted twice along the direction of the field lines). More specifically, the contraction $\mathcal{R}_{ab}n^an^b$ encodes the reaction of the field lines to their ``bending'' caused by the curvature of the host 3-space.\footnote{The magneto-curvature term in Eq.~(\ref{Ddu}) results from the vector nature of the $B$-field and the geometrical interpretation of gravity that Einstein's theory advocates. These two combine to guarantee a purely geometrical coupling between magnetism and spacetime curvature, which comes into play through the Ricci identities and goes beyond the standard interaction between matter and geometry seen in the Einstein equations. Among others, this coupling has been known to affect the propagation of electromagnetic radiation in curved spacetimes~\cite{NOO}-\cite{T3}.} Substituting the 3-divergence ${\rm D}^a\dot{u}_a$ from Eq.~(\ref{Ddu}) into Raychaudhuri's formula (see expression (\ref{tmRay1})), the latter reads
\begin{eqnarray}
\dot{\Theta}+ {1\over3}\,\Theta^2= -R_{ab}u^au^b+ c^2_a\,\mathcal{R}_{ab}n^an^b- 2\left(\sigma^2-\sigma^2_{_B}\right)+ 2\left(\omega^2-\omega^2_{_B}\right)+ \dot{u}^a\dot{u}_a\,. \label{MHDtmRay1}
\end{eqnarray}
with $R_{ab}$ and $\mathcal{R}_{ab}$ given by (\ref{Rab}) and (\ref{cRab}) respectively. When the right-hand side of the above is negative, an initially contracting congruence of worldlines will focus at a point (i.e.~$\Theta\rightarrow-\infty$) within finite proper time~\cite{W,Po}. In the opposite case, however, caustic formation and the ultimate collapse to a singularity will be averted. Therefore, negative terms on the right-hand side of the above assist implosion, while positive ones act in the opposite way~\cite{W,Po}.\footnote{It is worth comparing Eq.~(\ref{MHDtmRay1}) to the Raychaudhuri formula used in standard singularity theorems. The latter consider the convergence and focusing of irrotational, timelike geodesics, in which case we have~\cite{W,Po}
\begin{equation}
\dot{\Theta}+ {1\over3}\,\Theta^2= -R_{ab}u^au^b- 2\sigma^2\,, \label{Ray}
\end{equation}
without vorticity and 4-acceleration contributions (since $\omega=0=\dot{u}_a$ by default). This means that only stresses that assist the collapse have been included to the right-hand side of Eq.~(\ref{Ray}). Moreover, the absence of any 4-acceleration terms in the above relation implies that all the matter fields involved are homogeneously distributed. In this respect, the possibility of avoiding the ultimate singularity by involving non-gravitational forces (e.g.~see~\cite{Nov}-\cite{KB} and scenario (b) here), does not contradict the aforementioned theorems.}

Given that $\dot{u}_a\dot{u}^a>0$ always, the 4-acceleration stress seen at the end of Eq.~(\ref{MHDtmRay1}) resists contraction at all times. The remaining stresses act (in pairs) against each other in a way that manifests the counterintuitive impact of the tension effects. Since, in all cases of realistic stellar collapse, $\mathcal{R}_{ab}$ takes positive values along every spatial direction (including that of the $B$-field), the magneto-curvature stress ($c^2_a\,\mathcal{R}_{ab}n^an^b$) supports against the gravitational pull of the local matter, which itself is driven by the 4-Ricci curvature term ($R_{ab}u^au^b$). Also, while kinematic-shear distortions ($\sigma$) assist the collapse, the magnetic shear ($\sigma_{_B}$) inhibits contraction. The magnetic vorticity ($\omega_{_B}$), on the other hand, helps implosion, whereas vorticity proper ($\omega$) does the opposite. In all three cases, the magnetic tension ensures that the field lines react and try to counterbalance the effects of curvature, shear and rotation, namely of the agents that caused their deformation in the first place.

The fate of the magnetised collapse, as monitored by Eq.~(\ref{MHDtmRay1}), depends on the balance between the counteracting stresses seen on the right-hand side of that expression. In extreme situations, like those occurring during the gravitational implosion of a magnetised compact star or/and a black hole, it sounds plausible to argue that the ultimate fate of the collapse will be decided by the balance between the two curvature-related stresses on the right-hand side of (\ref{MHDtmRay1}). After all, these are the only purely general relativistic terms, since the rest will appear in flat Euclidean spaces as well (with the covariant derivatives replaced by partial ones). It also makes physical sense to claim that gravity (i.e.~curvature in Einstein's interpretation) should dominate the final stages of the collapse. Then, to leading order, Eq.~(\ref{MHDtmRay1}) reduces to
\begin{equation}
\dot{\Theta}+ {1\over3}\,\Theta^2= -R_{ab}u^au^b+ c^2_a\,\mathcal{R}_{ab}n^an^b\,. \label{MHDtmRay2}
\end{equation}
In the above one can clearly see the competing effects of all the matter fields driving the implosion on the one hand, and of the magnetic tension that acts against contraction on the other. Recall that $R_{ab}u^au^b=\kappa(\rho+3p)/2>0$ for all types of conventional matter and that the scalar $\mathcal{R}_{ab}n^an^b$ takes positive values in all realistic collapse scenarios. As we mentioned earlier, this magneto-curvature tension stress also carries the reaction of the field lines to distortions caused by the curvature of the host space.

What is most intriguing is that, as the collapse proceeds, the resistance of the magneto-curvature tension stress seen in Eq.~(\ref{MHDtmRay2}) increases. This is physically explained by the fact that the $B$-field is frozen into the collapsing matter and by the elasticity of the magnetic forcelines. More specifically, in realistic stellar collapse scenarios, the curvature of the imploding 3-space increases along all directions, including that of the $B$-field. This means that the scalar $\mathcal{R}_{ab}n^an^b$ becomes progressively more positive as the collapse progresses and also that the curvature induced magnetic-line distortion increases too. As a result, the resistance of the ``elastic'' magneto-curvature tension stress in Eq.~(\ref{MHDtmRay2}) grows stronger with the collapse. Therefore, although the aforementioned tension stress is triggered by the magnetic presence, its effect is driven by the ever increasing curvature of the imploding space. This makes it possible for relatively weak magnetic fields (i.e.~those with $c_a^2<1$) to trigger disproportionately strong magneto-curvature tension stresses in strong-gravity environments. In fact, as the 3-curvature diverges (i.e.~as $\mathcal{R}_{ab}n^an^b\rightarrow\infty$) during the final stages of the collapse, the resisting effect appears to become arbitrarily strong as well, even in the presence of a relatively weak $B$-field. Intuitively speaking, it appears as though the elasticity of the magnetic forcelines has been injected into the fabric of the host space, which seems to backreact against its own implosion. This entirely counterintuitive behaviour is what distinguishes the magneto-curvature tension stress in Eq.~(\ref{MHDtmRay2}) from other supporting stresses (like those due to pressure gradients, or to the vorticity, for example) and makes it unique.

With this in mind, it is worth going back to Melvin's 1965 paper~\cite{M2}, where he poses the following two questions: ``\textit{As the matter falls inward and carries the magnetic field with it, does this magnetic field inhibit the collapse? Does the magnetic field give rise to a counterpressure of decisive importance to the dynamics of collapse?}''. Half a century later, our work points towards positive answers to both of these questions. We found that magnetic fields exhibit a generic tendency to resist gravitational contraction, by naturally giving rise to (purely general relativistic) curvature-related supporting stresses that grow stronger as the collapse progresses. Moreover, by approaching the issue via a different route, we showed that the physical agent behind this magnetic behaviour is not the field's pressure but its tension, which makes the magnetised space act like an elastic spring under tension.

\section{Discussion}\label{sD}
\begin{figure}
\centering
\includegraphics[width=0.7\columnwidth]{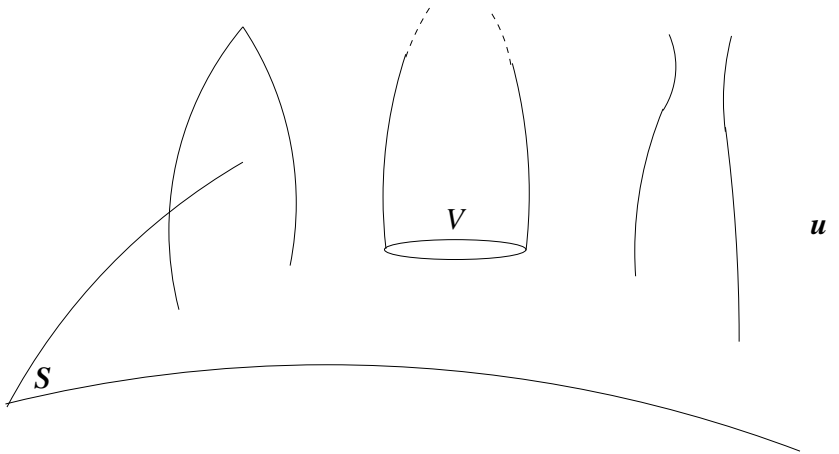}
\caption{Timelike worldlines of magnetised matter, with 4-velocity $u_a$ normal to the spatial 3-D hypersurface $S$ and an associated volume element $V$ (so that $\dot{V}/V=\Theta$). When gravity dominates and $R_{ab}u^au^b>c^2_a\,\mathcal{R}_{ab}n^an^b$, converging wordlines focus to form caustics within finite time (left). On the other hand, if the magneto-curvature tension stresses take over (with $c^2_a\,\mathcal{R}_{ab}n^an^b> R_{ab}u^au^b$), the collapse could proceed indefinitely and focusing may never happen (middle). There is also the possibility of a bounce, provided $\dot{\Theta}$ in Eq.~(\ref{MHDtmRay2}) turns positive (right).}  \label{fig:Wlines}
\end{figure}

Gravitational collapse does not only lead to compact stars and black holes, but ultimately it is also expected to cause the formation of spacetime singularities. Nevertheless, although the latter appear inevitable when dealing with geodesics, they could be averted when non-gravitational forces (and therefore non-geodesics) are involved.  Recall that repulsive Coulomb forces can in principle prevent the ultimate collapse of charged matter to a singularity~\cite{Nov}-\cite{KB}. Is it then conceivable that magnetism could be nature's way of preventing the catastrophic fate of a singularity? Following (\ref{MHDtmRay2}), it seems plausible that worldline focusing could be averted if the (tension induced) 3-Ricci stress outgrows its 4-Ricci counterpart (see Fig.~\ref{fig:Wlines}), namely when
\begin{equation}
c_a^2\mathcal{R}_{ab}n^an^b> R_{ab}u^au^b\,.  \label{cont}
\end{equation}
At this stage, it is still uncertain whether this (tentative) condition/requirement is naturally fulfilled in physically realistic situations. In principle, it depends on which of the two stresses grows faster towards the singularity. With this in mind, it sounds promising that the resisting magneto-curvature stress carries an additional direct geometrical (i.e.~gravitational) input. The latter comes from the Weyl field and more specifically from the electric component of the Weyl tensor (i.e.~from $E_{ab}$ -- see Eq.~(\ref{cRab}) and comments immediately after). In other words, the magnetic lines also react to distortions caused by tidal forces and gravitational waves, both of which are expected to grow arbitrarily strong as the collapse reaches its final stages. Note that, as it is well known, the Weyl field does not explicitly contribute to $R_{ab}$ (see expression (\ref{Rab}) earlier) and therefore to the 4-Ricci stress that drives the implosion in Eq.~(\ref{MHDtmRay2}). It is then conceivable that this ``selective'' contribution of the Weyl field could prove significant, by ``tilting'' the balance in favour of the supporting stress in condition (\ref{cont}).

Having said that, we should also emphasise that avoiding the focussing of the  non-geodesics worldlines of the charged particles, does not automatically guarantee a non-singular spacetime. Indeed, the latter may still contain timelike (as well as null) geodesics that are incomplete and singular (see also footnote~6 here). In this respect, our study is in no conflict with the Hawking-Penrose singularity theorems, at least no more than the work of~\cite{Nov}-\cite{KB} is. Recall that, in the latter studies, the non-geodesic worldlines of the charged particles were prevented from focusing by repulsive Coulomb stresses.

Even if $R_{ab}u^au^b> c_a^2\mathcal{R}_{ab}n^an^b$ and gravity ultimately wins, $B$-fields still show a remarkable generic tendency to reduce (or perhaps in some cases eliminate) the gravitational effects. The physical reason responsible for this (purely general relativistic) effect is the field's tension. In particular, it is the magnetic tension that cancels out the $B$-field's energy contribution to the curvature of the 3-D space. This leaves the twice contracted 3-curvature tensor, namely the scalar $\mathcal{R}_{ab}n^an^b$, free of any magnetic input and therefore ensures that the field lines do not feel their own gravitational pull, no matter how strong the latter may be. In the absence of any other sources, this ensures that magnetic forcelines that were originally parallel will remain so, in agreement with Melvin~\cite{M1,M2}. During the realistic gravitational contraction of charged matter, on the other hand, the scalar $\mathcal{R}_{ab}n^an^b$ takes positive values, deforms the magnetic forcelines and gives rise to curvature-related tension stresses that support against the contraction of the magnetised medium. No other known source behaves this way. What is most intriguing is that, as the collapse progresses and the curvature distortion increases (seemingly unrestrained), the resistance of the magneto-curvature tension stresses also grows (potentially unbounded as well), even when the $B$-field remains weak relative to the matterial content. Metaphorically speaking, one might say that the magnetised space reacts to the pull of gravity and tries to bounce back, as if it has itself acquired the elastic properties of the field lines.\\

\noindent \textbf{Acknowledgments:} The authors would like to thank Naresh Dadhich and Yuan Ha for their comments. CGT was supported by the Hellenic Foundation for Research and Innovation (H.F.R.I.), under the ‘First Call for H.F.R.I. Research Projects to support Faculty members and Researchers and the procurement of high-cost research equipment Grant’ (Project Number: 789). PM acknowledges
support by the Foundation for European Education and Culture.

\end{document}